\documentstyle[times,pramana,epsf,floats]{ias}

\def\hGpc{~$h^{-1}$Gpc}
\def\frtoday{le\space\number\day\space\ifcase\month\or
  janvier\or f\'evrier\or mars\or avril\or mai\or juin\or
  juillet\or ao\^ut\or septembre\or octobre\or novembre\or 
  d\'ecembre\fi\space \number\year}
\def\ltapprox{\,\lower.6ex\hbox{$\buildrel <\over \sim$} \, }
\def\deg{\ifmmode^\circ\else$^\circ$\fi}    

\newcommand\joref[5]{#1, #5, {#2, }{#3, } #4}

\newcommand\epref[3]{#1, #3, #2}

\def\aanda{A\&A}            
\def\cqg{ClassQuantGra} 
\def\mnras{MNRAS}

\begin{document}
\mark{{Topology of the Universe}{Boud Roukema}}
\title{Topology of the Universe:  background and recent observational
 approaches}  

\author{Boudewijn F. Roukema}
\address{Inter-University Centre for Astronomy and Astrophysics, 
 Post Bag 4, Ganeshkhind, Pune, 411 007, India (boud@iucaa.ernet.in)}

\abstract{Is the Universe (a spatial section thereof) finite or infinite? 
Knowing the global geometry of a Friedmann-Lema\^{\i}tre (FL) universe 
requires knowing both its curvature and its topology. A flat or
hyperbolic (``open'') FL universe is {\em not} necessarily infinite
in volume. 

Multiply connected flat and hyperbolic models are, in 
general, as consistent with present observations on scales of 
1-20{\hGpc} as are the corresponding simply connected flat and hyperbolic
models. The methods of detecting multiply connected
models (MCM's) are presently in their pioneering phase of development
and the optimal observationally realistic strategy is probably yet
to be calculated. Constraints against MCM's 
on $\sim$1-4{\hGpc} scales have been
claimed, but relate more to inconsistent assumptions on perturbation
statistics rather than just to topology. 
Candidate 3-manifolds based on hypothesised multiply imaged objects
are being offered for observational refutation.

The theoretical and observational sides of this rapidly developing
subject have yet to make any serious contact, but the prospects of 
a significant detection in the coming decade may well propel the
two together.}

\maketitle
\section{Cosmic topology}

This workshop is on observational cosmology: how observations 
confront cosmological theory. Unfortunately, one of the fundamental
aspects of Friedmann-Lema\^{\i}tre 
models of the Universe is weak in theoretical predictions.
General relativity says nothing about how big the Universe should
be. It describes curvature, which divides up 
constant curvature 3-manifolds (``spaces'') into three classes
corresponding to the three possible signs of curvature.

For example, a canonical flat multiply connected model is
the hypertorus, $T^3,$ which can be thought of as a cube whose 
opposite faces are identified. This is a flat 3-manifold without
any edges or boundaries, but finite in volume.
A $T^3$ universe may be as small as
1{\hGpc} or as big as the horizon for the same values of 
$\Omega_0,$ $\lambda_0,$ $\Omega_b,$ $\sigma_8$ and 
$H_0$.\footnote{These parameters
are defined as usual. The first two correspond to 
$\Omega_{m}$ and $\Omega_\Lambda$ in the popular 
Peebles \protect\cite{Peeb93} notation.} Evolution in the
luminosity functions of galaxies and quasars, the star formation
rate history of the Universe, and similar observational 
quantities do not distinguish between the
different models. They do not constrain the size of the 
Universe. Although the ``curvature radius'' and $H_0$
have strong effects on the size of the {\em observable} 
Universe, i.e. on the horizon radius, they only have weak
effects on the size of the Universe itself.

How can the theory (that spatial sections are 3-manifolds) 
be confronted with observations? In short, by photons travelling 
many times across the Universe so that multiple topological images
are seen of single objects. In a multiply connected universe, 
objects (or regions of CMB plasma) 
would be seen several times in different directions and (in general)
at different redshifts. This would be something like gravitational
lensing, except that the whole Universe would be the lens and 
the angular and radial distance differences in multiply imaged 
objects would
be, in general, big fractions of $\pi$ and of the horizon radius
respectively,
as opposed to arcsecond and sub-parsec differences in the case
of gravitational lensing.

\section{Recommended reading}

Recent reviews of the different observational strategies 
include \cite{RB98,LR99}
(the latter also includes
a brief historical and mathematical background). 

A fuller review including
theoretical aspects
of cosmic topology and pre-1993 observational work
is that of \cite{LaLu95}, but due to
exponential growth in the subject, the 
number of published articles on the subject has roughly doubled
since then.

Proceedings of the 1997 Cleveland and 1998 Paris workshops 
on cosmic topology 
are available as \cite{Stark98} 
and \cite{BR99} respectively. 

Mathematical tools, particularly including a ``census'' of a few
thousand small compact hyperbolic 3-manifolds are available 
at \cite{WeeksSP}.

\section{A survival kit for the observer: jargon}

The minimum concepts and jargon that the workshop participant or
reader should retain from the above literature are probably:
\begin{list}{(\roman{enumi})}{\usecounter{enumi}}
\item {\em ``compact''} essentially means finite in spatial volume
\item to avoid confusion, the word ``open'' is 
dropped in favour of {\em ``hyperbolic'', ``negatively curved'',
``$\Omega_0 < 1$''} or {\em ``$k < 0$''}; and ``closed'' is
dropped in favour of {\em ``elliptic'', ``spherical'', 
``positively curved'', ``$\Omega_0 > 1$''} or {\em ``$k > 0$} 
(otherwise, compact hyperbolic models would be referred to as
closed open models \dots)
\item {\em ``geodesic''} generally means a geodesic
in 3-space, but at times is used to mean a geodesic in 3+1 space-time
\item the entire ({\em comoving} spatial section of the) 
Universe can be represented as a polyhedron embedded in 
$H^3,$ $R^3$ or $S^3$ (for $k < 0, =0, > 0$ respectively) of which 
faces are identified with one another in some way --- this is the
{\em ``fundamental polyhedron''} or {\em ``Dirichlet domain''}
\item by pasting together copies of the fundamental domain, an 
 space $H^3,$ $R^3$ or $S^3$ (respectively) 
can be constructed which corresponds, for the observer, to
the apparent space in which objects at high redshift are located
under the hypothesis of trivial topology\footnote{``Trivial topology'' 
refers here to the property of having 
a trivial $\pi_1$ homotopy group.} --- this is 
termed the {\em ``universal covering space'',} $\widetilde{M}$
\item in the covering space, the isometries mapping multiple
{\em ``topological images''} (or {\em ``topological clones''})
to one another form a group, $\Gamma$, whose elements are 
linear combinations of a set of {\em ``generators''}
\item the 3-manifold can formally be written as $M=\widetilde{M}/\Gamma$
\item for convenience, one often swaps thinking and calculating
between the fundamental polyhedron and the covering space.
\end{list}

\section{An example of a candidate 3-manifold}

In the commonly studied case of the rectilinear toroidal models,
multiple topological images of an object form a rectilinear grid
in comoving space. Among a small selection of the brightest known
galaxy clusters, three form a right angle of equal arm lengths
to within $2-3\deg$ and $1\%$ accuracy respectively 
\cite{RE97}. 

Are the Coma cluster, RX~J1347.5-1145 and CL~09104+4109 
three images of a single cluster or is the right angle just a coincidence? 

A list of arguments for and against this $T^2$ 
candidate is provided in the discussion
section of \cite{RBa99}, and a comparison with COBE data is
presented in \cite{Rouk99}.

\section{Projects}

This is a workshop. The following are ideas suggested for projects.

\subsection{Theory}

\begin{list}{(\roman{enumi})}{\usecounter{enumi}}
\item What should the topology of the Universe be? {\em Can a theory
of quantum gravity or of quantum cosmology make any serious
predictions about what the topology of the Universe should be
at $t\sim t_0$? }
\item A group $\Gamma$ relates the covering space 
$\widetilde{M}$ of a multiply
connected universe to the fundamental polyhedron 
$M=\widetilde{m}/\Gamma.$ The standard model of particle physics
relates different particles to one another by a group,
e.g. SU(2)$_L \times$ U(1)$_{YW} \times$ SU(3)$^C$. {\em Could the 
Universe be considered a particle at the quantum epoch 
and the spatial transformations of $\Gamma$ be related to 
the gauge bosons?}
\end{list}

\subsection{Observation}

\subsubsection{Methods}

\begin{list}{(\roman{enumi})}{\usecounter{enumi}}
\addtocounter{enumi}{2}
\item The classical magnitude-redshift relation yielded only
weak constraints on the curvature parameters ($\Omega_0,$ $\lambda_0$)
until an empirical way of improving supernovae of type Ia as 
standard candles was 
devised. The results are impressive, even though theoretical 
understanding of the method of sharpening the standard candle 
is weak \cite{SCP9812}.\footnote{Cosmic topology could provide 
high precision estimates of the curvature parameters. 
Detection of 5-10 multiple topological
images of an object up to $z \sim 2-3$ would be sufficient 
to estimate $\Omega_0$ and $\lambda_0$ to better than 1\% and
10\% respectively \protect\cite{RL99}.}
{\em Could some sort of similar ``trick'' improve the
presently published methods to the point of extracting a significant
topological detection?}
\item {\em Realistic simulations including all the observational 
difficulties
could be used to optimise the cosmic crystallography 
\cite{LLL96,LLU98,ULL99a} 
and local 
isometry search methods \cite{Rouk96,ULL99a}.}
\item {\em Realistic simulations and analysis 
should also be used to find the best
way to apply the matched circles principle 
\cite{Corn96,Corn98b,Weeks98,Rouk99}.}
\end{list}

\subsubsection{Candidates}

\begin{list}{(\roman{enumi})}{\usecounter{enumi}}
\addtocounter{enumi}{5}

\item {\em Generate specific candidates.}
\item {\em Observationally refute these in order to understand
systematic errors.}
\end{list}

\subsubsection{New catalogues}

\begin{list}{(\roman{enumi})}{\usecounter{enumi}}
\addtocounter{enumi}{7}

\item radio: GMRT --- $5\ltapprox z \ltapprox 10$ ? proto-clusters
\item mm/sub-mm: MMA/LSA --- $5\ltapprox z \ltapprox 10$ ? galaxies
\item cm: MAP, Planck ---$ z \approx 1100$ or \dots (integrated 
Sachs-Wolfe 
effect) $z \ll 1100$ CMB (plasma); $z \sim 1 - 3$ clusters (SZ effect)
\item optical: SDSS, VLT --- $ z \sim 1 - 3$ ? quasars, galaxies
\item Xray: XMM --- $z \sim 1 - 2$ clusters, quasars
\end{list}

\subsubsection{Local (10kpc $-$ 100Mpc)}

\begin{list}{(\roman{enumi})}{\usecounter{enumi}}
\addtocounter{enumi}{12}
\item {\em Understand the Galaxy (or the local unit of 
large scale structure)
well enough to say what its topological image must have looked like
at $z\sim 2-5$ and from an ``arbitrary'' angle.} This would be a ``safe''
theme for a thesis project, since the theoretical and/or observational
work done in understanding the Galaxy would be valid independently of
its use in identifying or refuting high redshift topological images
of the Galaxy.
\end{list}


\end{document}